%% file: Main.tex
\documentclass[a4paper, 9pt, twocolumn]{article}
\usepackage{geometry}                
\usepackage{graphicx}
\usepackage{amssymb}
\usepackage{amsthm}
\usepackage{epstopdf}
\usepackage{mathtools}
\usepackage[shortlabels]{enumitem}
\usepackage{dsfont}
\usepackage{cuted}
\usepackage{sidecap}
\usepackage[usenames,dvipsnames]{color}
\usepackage[thinspace,mediumqspace,Grey,squaren]{SIunits}
\usepackage{bbm}
\usepackage[T1]{fontenc}
\usepackage[utf8]{inputenc}
\usepackage{authblk}
\usepackage[font=small,labelfont=bf ,textfont={small, sf}]{caption}
\usepackage{titlesec}
\usepackage{url}
\usepackage{hyperref}
\titleformat{\subsection}[runin]{\normalfont\bfseries}{\thesubsection.}{3pt}{}

\input{Definitions}

\begin{document}

\title{Pooling single-cell recordings: Scalable inference through heterogeneous kinetics}

\author[1]{Christoph Zechner}
\author[1, 2]{Michael Unger}
\author[2, 3]{Serge Pelet}
\author[2]{Matthias Peter}
\author[1,4]{Heinz Koeppl\thanks{corresponding author: koeppl@ethz.ch}}
\affil[1]{Automatic Control Laboratory, ETH Zurich, Switzerland}
\affil[2]{Institute of Biochemistry, ETH Zurich, Switzerland}
\affil[3]{Department of Fundamental Microbiology, University of Lausanne, Switzerland}
\affil[4]{IBM Research Laboratory - Zurich, Switzerland }
\renewcommand\Authands{ and }

\date{}

\maketitle

\newpage{}

{\noindent \sf \fontsize{9}{0} \selectfont \textbf{
Mathematical methods together with measurements of single-cell dynamics provide unprecedented means to reconstruct intracellular processes that are only partly or indirectly accessible experimentally.  
To obtain reliable reconstructions the pooling of measurements from several cells of a clonal population is mandatory. The population's considerable cell-to-cell variability originating from diverse sources poses novel computational challenges for process reconstruction. 
We introduce an exact Bayesian inference framework that properly accounts for the population heterogeneity but also retains scalability with respect to the number of pooled cells. The key ingredient is a stochastic process that captures the heterogeneous kinetics of a population. The method allows to infer inaccessible molecular states, kinetic parameters, compute Bayes factors and to dissect intrinsic, extrinsic and technical contributions to the variability in the data. We also show how additional single-cell readouts such as morphological features can be included into the analysis. We then reconstruct the expression dynamics of a gene under an inducible GAL1 promoter in yeast from time-lapse microscopy data. Based on Bayesian model selection the data yields no evidence of a refractory period for this promoter.
}}

\section*{Introduction}

Statistical inference of unobserved molecular states and unknown parameters that characterize an intracellular process is instrumental for the advance of quantitative biology. In particular, single cell assays provide an unprecedented source of data to perform this inference. They provide access to the possibly significant stochastic nature of those processes but also expose the considerable heterogeneity in clonal populations of cells. 

From the viewpoint of inference, two classes of single-cell data may be distinguished. The first are population snapshot data, provided for instance by cytometry techniques \cite{Zechner2012, Hasenauer2011,Ornatsky2010} or FISH (fluorescence-in-situ-hybridization, \cite{Raj2006, munsky2009listening}, where -- when measured over time -- any temporal correlation of a cell's dynamics is necessarily lost. The second class are time-lapse single-cell data \cite{Harper2011, Suter2011}, where the dynamics of individual cells can be followed over several time points. Such a recording can be considered a partially observed sample path of the underlying process and thus, not only gives us a handle on the marginal distribution at each time point but also on how the temporal correlation of the process looks like. Evidently, such information is extremely helpful for inference and renders live-cell measurements superior in this respect -- taking aside its limitation in terms of throughput compared to flow cytometry, for instance.  

Recently, we proposed a complete inference framework for single-cell snapshot data \cite{Zechner2012}. In particular, cell-to-cell variability was mathematically accounted for and time-lapse flow-cytometry data was exemplarily used to infer kinetic parameters of a gene expression system in yeast. In this work we lay out a corresponding inference scheme for time-lapse live-cell data. A very different approach needs to be followed here in order to account for the sample path nature of the acquired data.

Several inference techniques for stochastic chemical kinetics based on single-cell time-lapse data have been proposed recently \cite{Amrein2012, Toni2009, Golightly2011, Opper2007, Girolami2012}. Focus was put on the inference from the noisy observation of a single sample path, because the extension to multiple sample paths is computationally straightforward if one assumes no heterogeneity. In practice, pooling of several single-cell recordings is necessary to obtain reasonable estimates, which in turn generates difficulties due to the extrinsic contribution to the observed heterogeneity of a population \cite{Elowitz2002, Colman-Lerner2005, Snijder2011}. Examples of such extrinsic factors are cell-cycle stage or translation efficiency of the cells. The effect of those factors on the dynamics of the actual process under study has been investigated in \cite{Bowsher2012, Hilfinger2011}. First attempts have been made in \cite{Koeppl2012, Hasenauer2011} to devise inference techniques that account for and estimate such heterogeneity. 

Imaging-based single-cell techniques can additionally capture morphological features of cells such as their volume or shape. While those were shown to correlate well with extrinsic factors \cite{Rinott2011, Snijder2011}, previous inference approaches do not offer a principled way to exploit such readouts.

By addressing extrinsic variability, the present work goes significantly beyond previous inference efforts. Based on the innovation theorem for counting processes \cite{Aalen1978}, we obtain a marginal stochastic process by integrating the heterogeneous Markov process over the extrinsic factors as well as fixed kinetic parameters with prior uncertainty. This results in a scalable inference scheme, the parameter dimension of which is independent of the population size. In conjunction with the time-lapse microscopy measurements, the method allows us to reconstruct the dynamics of an inducible gene expression system in yeast from pooled time-lapse microscopy data. 

\section*{Results}

\subsection*{Mixed-effect Models and Measurements.}
We consider a continous-time Markov chain (CTMC) describing the time-evolution of a well-mixed reaction system with $\StateDim$ chemical species and $\ReactionDim$ reaction channels. The random state vector $\State(t)$ collects the copy numbers of species at time $t$ and random paths thereof on the time interval $[0, t]$ are denoted $\StateInt$. 
Subsequently, we will
use the convention to denote a random quantity by an upper case letter and its realization by the corresponding lower case letter. The propensity functions of the reactions are assumed to be of the form $h_{j}(\state) = \kparameter_{j} g_{j}(\state)$ with $\kparameter_{j}$ the stochastic rate constant and $g_{j}$ a function of the current state $\state$. Moreover, we denote the set of all stochastic rate constants as $\kParameters = ( \kParameters_{1}, \ldots, \kParameters_{\ReactionDim} )$. 
Following a Bayesian approach, the random parameters $\kParameters$ get equipped with a prior believe through a probability distribution $p(\kparameters)$.   
Note that besides the reaction rate constants, a single-cell stochastic model of a cellular process may additionally be parametrized in terms of its initial conditions, size of the reaction compartment, the state of the molecular environment in which the process is embedded and so forth. We denote such a complete parametrization of the Markov chain as $\AllParameters$.

Apart from intrinsic fluctuations of the stochastic process, extrinsic variations will render process dynamics different across cells in a population even in the absence of intrinsic fluctuations -- suggesting the use of individual Markov chains for each cell. Hence, we allow some of the process parameters to vary across cells. We denote those as {\em extrinsic factors} 
%
$\ExtrinsicVariable \subseteq \AllParameters$. They can either be assumed constant \cite{Zechner2012, Hasenauer2011} or dynamically changing \cite{Hilfinger2011,Bowsher2012}. Extrinsic factors include rate constants of aggregated multi-step reactions that dependent on the presence of intermediates (e.g. translation rate), initial conditions or similarly total copy numbers of conserved proteins, ribosome numbers, cell volume or cell-cycle stage - to name just a few. Here, we take a macroscopic view and address the evident heterogeneity in a cell population that is believed to roughly stay unchanged during the course of an experiment. In turn, we neglect the microscopic picture \cite{Hilfinger2011,Bowsher2012} that some extrinsic factors, on top of being different across cells, also fluctuate temporally and modulate the considered process. For instance, ribosomes are themselves subject to intrinsic fluctuations.  Hence, we assume that those fluctuations are small compared to the constant offsets across cells and to the intrinsic fluctuations of the actual process under study. On the other hand, some extrinsic factors, such as the cell-cycle stage may not fluctuate but rather vary deterministically and slowly with respect to the experimental time span \cite{Dunlop2008, Shahrezaei2008, Rausenberger2008}, such that quasi-steady state arguments can support a time-invariance assumption.  

In statistics the resulting heterogeneous population models are known as \textit{mixed-effect} models, where the complement subset $\SharedParameters = \AllParameters\setminus\ExtrinsicVariable$ are {\it shared} among cells and correspond to the \textit{fixed} effects, whereas the extrinsic factors refer to the \textit{random} effects. Examples in $\SharedParameters$ are rate constants of elementary reactions, such as protein associations that are determined by the biophysics of the two proteins and should not vary across cells. Following the time-invariance of extrinsic factors $\ExtrinsicVariable$, we assign to them a probability distribution $p(\extrinsicVariable \mid \hyperParameters)$, where the {\it extrinsic statistics} $\hyperParameters$ is a set of parameters specifying the shape of this distribution and hence the population's heterogeneity. Given a population of $\CellCount$ cells, the $m$-th cell's state is then described by a conditional Markov chain $\MarkovChain^{m} \mid (\ExtrinsicVariable^{m}, \SharedParameters)$.


Experimentally we can retrieve a corrupted abundance measure for a small subset of molecular species for $\CellCount$ different cells at each measurement time $t_l$. 
The associated acquisition error is characterized by a conditional measurement density 
$p(\measurementTime^{m} \mid \state_{l}^{m}, \measurementParameters)$, with $\state^m_{l} \equiv \state^m(t_l)$ and $\measurementParameters$ an unknown distribution parameter such as the acquisition noise variance.  Furthermore, we define the state trajectory of cell $m$ between the $l$-th and the $k$-th measurement time as $\StateInt_{l:k}^{m}$ and denote by $\MeasurementInt_{l:k}^{m}$ the corresponding set of measurements\footnote{Note that in contrast to $\MeasurementInt_{l:k}^{m}$, the state trajectory $\StateInt_{l:k}^{m}$ denotes a continuous-time sample path between $t_{l}$ and $t_{k}$.}. 
Morphological features are accounted for by introducing {\it morphological covariates} $\Covariates^{m}$ and hypothesizing a statistical dependency between these covariates and the extrinsic factors $\ExtrinsicVariable^{m}$. This dependency is described by a conditional density 
$p(\covariates^{m} \mid \extrinsicVariable^{m}, \morphParameters)$, 
with $\morphParameters$ a set of shape parameters characterizing this conditional density. 
A schematic representation of the resulting mixed-effect Markov model is given in Figure 1a and the corresponding hierarchical Bayesian network is depicted in Figure 1b.

\begin{figure*}
	\centering
	\includegraphics[width=0.99\textwidth]{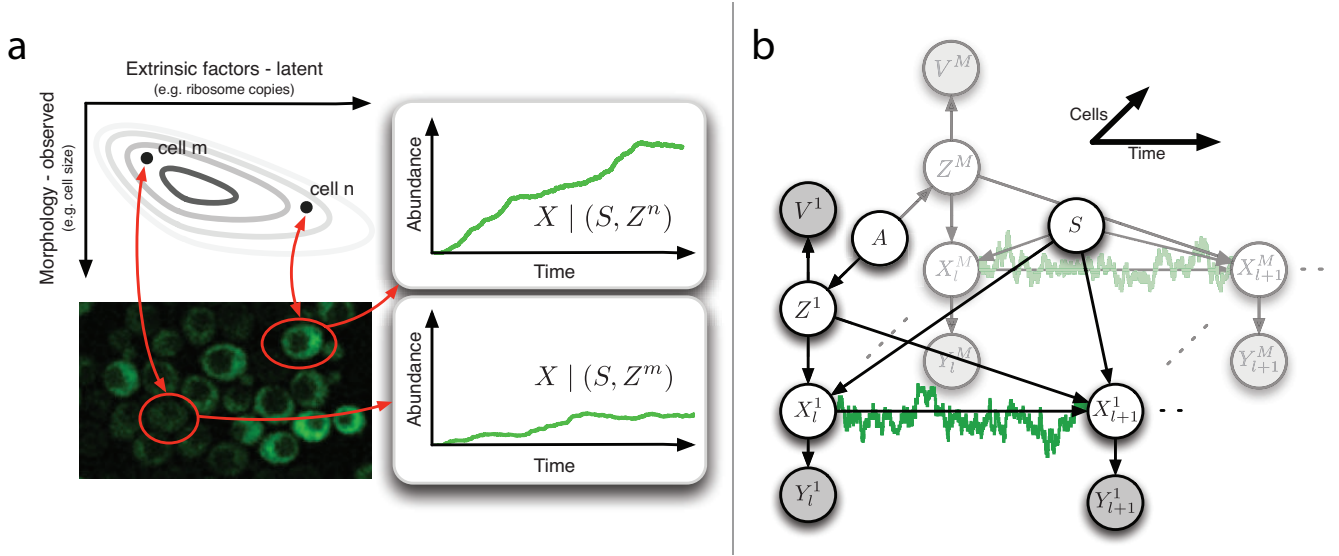}
	\caption{Modeling heterogeneous microscopy data.
(a) Schematic generative model of the experimental data. On top of intrinsic fluctuations, extrinsic factors and their morphological covariates render individual cells different. Recorded intensitiy levels are hence characterized by a parameter set $\SharedParameters$ that is shared across cells and a set of individual (i.e. extrinsic) parameters $\ExtrinsicVariable^i$.  
(b) Corresponding Bayesian mixed-effect model. Nodes denote random variables and statistical dependency (and causality) is indicated by directed edges. Gray-shaded nodes correspond to experimentally accessible quantities, whereas all other nodes refer to latent variables. Extrinsic factors $\ExtrinsicVariable^i$ are drawn from a common distribution parametrized by $\HyperParameters$.  Note that for clarity, the nuisance parameters $\MeasurementParameters$ and $\MorphParameters$ defined in the text are not shown. 
	}
		\label{fig:Figure1}
\end{figure*}

\subsection*{Heterogeneous Kinetics.}
According to the above mixed-effect model, the dimension of the parameter space, on which we aim to perform inference subsequently, is scaling with the number of considered cells $\CellCount$. Hence, a marginalized process model, where the unknown extrinsic factors are integrated out, appears critical to achieve practical tractability. Formally, the dynamic state of each cell $m$ is then described by one and the same \textit{marginal process} $\MarkovChain^{m} \mid ( \SharedParameters, \HyperParameters) := \MarkovChain  \mid ( \SharedParameters, \HyperParameters)$ with path density $p( \stateInt  \mid  \sharedParameters,\hyperParameters) = \int p( \stateInt \mid\extrinsicVariable, \sharedParameters)p(\extrinsicVariable \mid \hyperParameters) \df \extrinsicVariable$. For the sake of illustration, let us assume $\AllParameters = \kParameters$ and $\ExtrinsicVariable$ to be one-dimensional, namely the stochastic rate constant of reaction $j$ that hence randomly varies across the population. 
Based on the innovation theorem for counting processes \cite{Aalen1978}, we show that the marginalized dynamics again follows a jump process, where the propensity corresponding to the extrinsic parameter can be generally written as
\begin{equation}
	\MargProp_{j}(\stateInt,t) = \Expect{\ExtrinsicVariable \mid \stateInt, \hyperParameters} g_{j}(\state(t)),
	\label{eq:MarginalHazardGeneral}
\end{equation}
with $\Expect{\ExtrinsicVariable \mid \stateInt, \hyperParameters}$ as the conditional expectation of $\ExtrinsicVariable$ given a complete sample path $\stateInt$ and the extrinsic statistics $\hyperParameters$. For instance, if $\ExtrinsicVariable$ follows a Gamma distribution $p(\extrinsicVariable \mid \hyperParameters) = \GammaDist(\alpha, \beta)$, with $\hyperParameters = (\alpha,\beta)$ we have that  $p(\extrinsicVariable \mid \stateInt, \hyperParameters)  = \GammaDist(\alpha + r_{j}, \beta + \int_{0}^{t} g_{j}(\state(\tau)) \df\tau )$ \cite{Wilkinson2006,Kuechler1997} and hence,
\begin{equation}
	\MargProp_{j}(\stateInt,t) = \frac{\alpha + r_{j}}{\beta + \int_{0}^{t} g_{j}(\state(\tau)) \df\tau} g_{j}(\state(t)),
	\label{eq:MarginalHazard}
\end{equation}
where $r_{j}$ denotes the number of occurrences of reaction $j$ in $\stateInt$. For CTMCs, (\ref{eq:MarginalHazardGeneral}) can be rewritten \cite{Aalen1987} in terms of the moment-generating function of $\ExtrinsicVariable \mid \HyperParameters$; mathematical proofs and derivations for the general case of multivariate extrinsic variables are provided in section S.1 in the Supporting Information. 
If morphological covariates are included into the analysis, the marginalization can be performed analogously by constructing a marginal-conditional process $\MarkovChain^{m}  \mid (\SharedParameters, \HyperParameters, \MorphParameters, \Covariates^{m})$ with path density $p( \stateInt^{m}  \mid  \sharedParameters, \hyperParameters, \morphParameters, \covariates^{m})$ (see section S.1 in the Supporting Information). 
With the above marginalization, the randomness over extrinsic factors is exposed by the stochastic process itself which now shows a direct dependency on the extrinsic statistics $\HyperParameters$. Note that as a consequence, the Markov property is lost and moreover, the propensity $\MargProp_{j}(\stateInt,t)$ becomes explicitly time-varying. However, when augmenting the state space by the summary statistics $T(\stateInt) = (\, r_{j},\, \int_{0}^{t} g_{j}(\state(\tau)) \df\tau\,)$ of the path $\stateInt$, the Markov property is recovered and hence, stochastic simulation can efficiently be performed using available methods \cite{Anderson2007}, see section S.1.1 in the Supporting Information). 
On top of several computational advantages, the marginal process provides an important means to properly describe stochastic biomolecular dynamics in presence of extrinsic variability. More specifically, it reflects the dynamics of a cell with \textit{unknown} extrinsic factors and hence, coincides with typical real-world scenarios. In contrast to traditional stochastic models that rely on full parametric knowledge of the considered process, the marginal jump process can take into account uncertainty over extrinsic factors and thus, provides a principled framework to study kinetics in a heterogeneous population.

\subsection*{Statistical Inference.}
Before we outline the inference procedure in more detail, we discuss an important implication of the above marginalization approach. 
In particular, it follows that any kind of parametric uncertainty, linearly entering the propensities can be integrated out and directly encoded into the stochastic process. This seems particularly useful from a Bayesian viewpoint, where parameters are characterized by prior uncertainty. For instance, if the $i$-th kinetic parameter $\kParameters_{i}$ is associated with a known prior distribution, e.g. $p(\kparameters_{i}) = \GammaDist(\alpha_{i}, \beta_{i})$, the marginal propensities are obtained in analogy to (\ref{eq:MarginalHazard}). Marginalization with respect to every kinetic parameter in conjunction with a Bayesian filtering approach leads to the following inference scheme, which we refer to as \textit{dynamic prior propagation} (DPP). 
The goal of the scheme is to compute the marginal posterior distribution

\begin{strip}
\rule{\textwidth}{0.4pt}
\begin{equation}
	\begin{split}
	p\left(\stateInt_{1:N}^{1}, \ldots, \stateInt_{1:N}^{M}, \hyperParameters, \morphParameters, \measurementParameters \mid ( \measurementInt_{1:N}^{1}, \covariates^{1} ), \ldots, ( \measurementInt_{1:N}^{M}, \covariates^{M})\right) &\propto \Bigg[ \prod_{m=1}^{M} \left( \prod_{l=1}^{\TimeIndex} p(\measurementTime^{m} \mid \stateDiscreteTime^{m}, \measurementParameters) \right) 	p(\stateInt_{1:N}^{m} \mid \hyperParameters, \morphParameters, \covariates^{m}) p(\covariates^{m} \mid \morphParameters, \hyperParameters) \Bigg]  \\
	&\times p(\hyperParameters) p(\morphParameters) p(\measurementParameters),
	\end{split}
	\label{eq:FullPosterior}
\end{equation}
\rule{\textwidth}{0.4pt}
\end{strip}

where $(\measurementInt_{1:N}^{m}, \covariates^{m})$ denotes the tuple of measurements available for the $m$-th cell. 
In principle, a Markov chain Monte Carlo (MCMC) scheme can be applied directly to sample from (\ref{eq:FullPosterior}). However, by the curse of dimensionality such approaches are prohibitive in terms of acceptance ratios when considering multi-dimensional dynamics over multiple measurement time points. According to a Bayesian filtering approach, the posterior distribution is determined by applying Bayes' law recursively over time. Consequently, the original sampling problem breaks up into a sequence of sub-problems with significantly reduced dimensionality. The posterior distribution at time $t_{l}$ is then given by

\begin{strip}
\rule{\textwidth}{0.4pt}
\begin{equation}
	\begin{split}
	p\left(\stateInt_{1:l}^{1}, \ldots, \stateInt_{1:l}^{M}, \hyperParameters, \morphParameters, \measurementParameters \mid ( \measurementInt_{1:l}^{1}, \covariates^{1}), \ldots, (\measurementInt_{1:l}^{M}, \covariates^{M}) \right) &\propto \left[ \prod_{m=1}^{M} p(\measurementTime^{m} \mid \state_{l}^{m}, \measurementParameters)  
	p\left(\stateInt_{l-1:l}^{m} \mid \state_{l-1}^{m}, \hyperParameters, \morphParameters, \covariates^{m}, T(\stateInt_{1:l-1}^{m}) \right) \right] \\
	& \times p\left(\stateInt_{1:l-1}^{1}, \ldots, \stateInt_{1:l-1}^{M}, \hyperParameters, \morphParameters, \measurementParameters \mid ( \measurementInt_{1:l-1}^{1}, \covariates^{1}), \ldots, (\measurementInt_{1:l-1}^{M}, \covariates^{M}) \right).
	\end{split}
	\label{eq:BayesianRecursion}
\end{equation}
\rule{\textwidth}{0.4pt}
\end{strip}

Although analytically intractable, we can sample from (\ref{eq:BayesianRecursion}) using a Metropolis-Hastings sampler, for instance.
Recursive sampling approaches inherently suffer from sample degeneracy as soon as constant parameters are estimated in addition to the dynamic states. While a majority of the parameters is integrated out thanks to the marginal process framework, $\HyperParameters$, $\MorphParameters$ and $\Omega$ remain in the model. A standard approach to avoid such degeneracies is to apply an invariant kernel to the static parameters at each time instance, such as to maintain diversity among parameter samples \cite{Storvik2002}. Here, we use a Metropolis-within-Gibbs MCMC scheme, where the latent space is further divided into blocks which are successively resampled. This requires sampling from the full conditional distributions, which can be determined using the notion of Markov blankets \cite{Koller2009}. A full description of the algorithm can be found in section S.2 in the Supporting Information. 
Note that posterior distributions over $\ExtrinsicVariable^{1}, \ldots, \ExtrinsicVariable^{M}$ and $\SharedParameters$ are not directly computed by the marginalized inference scheme. However, they can be easily reconstructed via the law of conditional probability, such as described in section S.2.4 in the Supporting Information.

\subsection*{Synthetic Data: Two-State Gene Expression Model.}
We first studied the proposed inference framework using synthetic data of a simple two-state gene expression model \cite{Raj2006} given in Figure 2a under realistic measurement conditions. We assume that the target gene can be activated upon application of an exogeneous signal -- for instance via stress-induced translocation of certain transcription factors. 
Extrinsic variability was simulated by introducing a Gamma-distributed variability in the translation rate $\kparameter_{5}$. Data was collected for $M=20$ cells, on which we applied DPP using 10000 samples per time instance (for further details, see caption of Figure 2a and Figure 2b). The inferred posterior distributions of the kinetic parameters, the extrinsic statistics as well as the acquisition noise parameter are depicted in Figure 2b (for a density plot of the morphological shape parameters see Figure S.5 in the Supporting Information). Furthermore, state inference with respect to mRNA and protein levels is demonstrated in Figure 2c using two exemplary cells with different translation efficiencies. 

\begin{figure}
	\centering
	\includegraphics{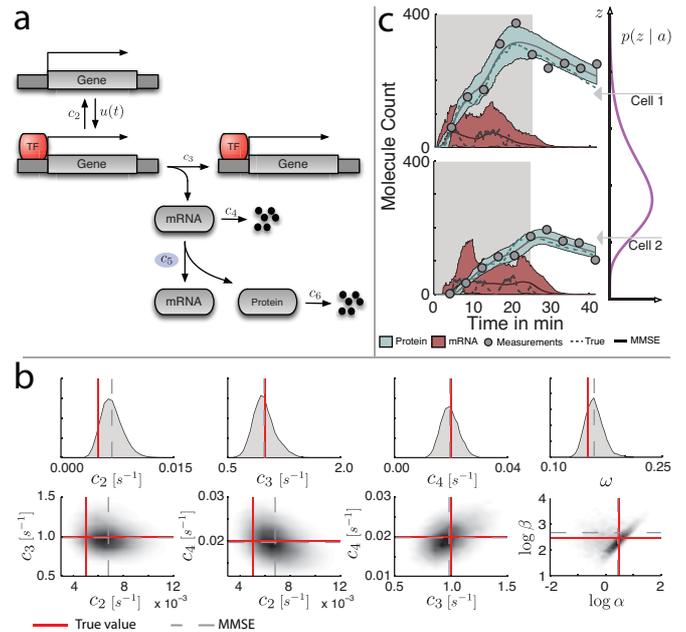}
	\caption{Parameter and state inference using synthetic data. 
	(a) Two-state gene expression model. The gene activation event is controlled by a time-varying rate $u(t)$ and subsumes transcription initiation events such as recruitment of transcription factors, RNA polymerase and possibly chromatin remodeling. Upon activation of the gene, mRNA can be transcribed to further yield new proteins. All reactions are modeled according to mass-action. Altogether, the model comprises four species and six reactions, whereas we assume a Gamma-distributed heterogeneity in the translation efficiency (i.e., $\kparameter_{5}$) with distribution parameters $\alpha$ and $\beta$. We assume that $\TimeIndex=10$ noisy (log-normally distributed with unknown scaling parameter) measurements of the protein abundance can be obtained at equally spaced time points within a total interval of roughly 40 minutes.
	(b) Parameter inference from protein time series using $M=20$ cells. Inference results are shown for the three kinetic parameters ($c_{2}$, $c_{3}$ and $c_{4}$), the extrinsic statistics $\alpha$ and $\beta$ (describing the heterogeneity over $c_{5}$) and the scaling parameter $\omega$ of the acquisition noise. The sequential MCMC scheme was performed using 10000 samples per time instance.
	(c) Inferred mRNA and protein abundance. The 5\%- and 95\%-quantiles of the inferred state distributions as well as their mean values, i.e., minimum mean square error (MMSE) estimates are shown for two representative cells with different translation efficiencies; the gray-shaded area indicates the window of induction. The thick violet line illustrates the assumed Gamma distribution over the extrinsic factor, i.e., the translation rate. }
		\label{fig:Figure2}
\end{figure}

Importantly, the inferred posterior distribution over the latent states can be used to reconstruct switching events of the target gene. We remark that a simple mode detection on the marginal distribution (time-point-wise) does not yield a valid sample path and hence cannot be used to extract timing statistics, such as gene on/off times.   
Consequently, we determined the most likely posterior (MAP) switching path of the gene from which timing statistics may be recovered. This approach to reconstruct switching events is exact and contrasts other approximate schemes \cite{Harper2011}, the validity of which was not demonstrated on synthetic data where the actual switching path is known. In general, the inverse problem of reconstructing a gene switching path from the slow protein dynamics is considerably ill-posed and we expect the posterior density over the switching paths to be close to degenerate. 
However, for the simulation study, Figure 3 indicates that accurate detection of the gene state is indeed possible within the realistic scenario considered here. For this test we simulated a pulsed induction of gene expression.    

\begin{figure}
	\centering
	\includegraphics{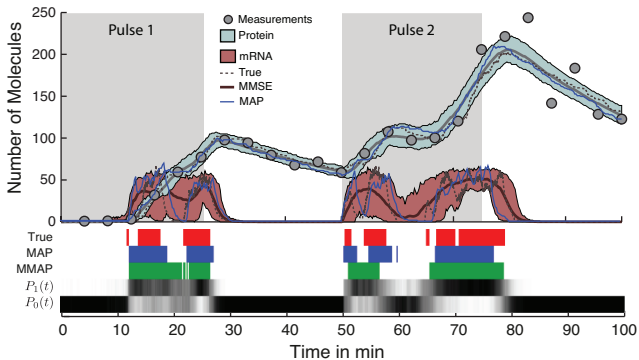}
	\caption{Reconstructed gene expression dynamics for a double-pulsed induction from noisy protein measurements (circles). Shaded areas denote 5\%- and 95\%-quantiles of the posterior distributions over states and the true values (solid), the MMSE estimates (dashed) and the maximum a posteriori (MAP) paths (blue) are shown. Also shown are the posterior probabilities for the gene to be inactive or active, $P_{0}(t)$ and $P_{1}(t)$, respectively; black and white coloring corresponds to probability one and zero, respectively.  Furthermore, we computed the marginal MAP (MMAP) state estimate (green) based on $P_{0}(t)$ and $P_{1}(t)$. 
			}
		\label{fig:Figure2B}
\end{figure}

\subsection*{Experimental Data: $\beta$-Estradiol-Induced Gene Expression.} The DPP-based inference was used to reconstruct the expression dynamics of an artificially controlled gene expression system in yeast. A widely used system to control the expression of genes under a GAL1 promoter in \textit{Saccharomyces cerevisiae} is based on the hormone-dependent activation of the chimeric transcription factor GAL4DBD.ER.VP16 (GEV) \cite{Louvion1993, McIsaac2011}. GEV consists of a strong transcriptional activator, made by fusing the GAL4 DNA binding domain (GAL4DBD) with the hormone-binding domain of the human estrogen receptor (ER) \cite{Louvion1993} and the transcription activating domain of the herpes simplex virus protein VP16 \cite{Sadowski1988}. In its inactive state, GEV associates with the Hsp90 chaperone complex and resides in the cytoplasm. Upon addition of the exogenous hormone $\beta$-estradiol to the extracellular medium, $\beta$-estradiol diffuses through the cell membrane and binds to the GEV's ER. Thereby, Hsp90 disassociates from the complex and active GEV translocates to the nucleus where it's GAL4DBD recognizes and binds to GAL promoter regions. VP16 then activates transcription of the downstream gene.

We engineered a strain that allows a combined readout of GEV translocation and $\beta$-estradiol-induced gene expression. A fully functional GEV-mCherry construct in combination with a nuclear marker allows to compute a ratio of nuclear to cytoplasmic GEV, serving as our model input. In the same strain, a destabilized \cite{Varshavsky1996} version of the Venus fluorescent protein (Y-Venus) was placed under control of a GAL1 promoter, which allows a more accurate tracking of the gene expression dynamics. For a detailed description of the strain see S.4.1 in the Supporting Information.


The fluorescence microscopy experiments were performed using a flow chamber that allowed us to rapidly exchange the extracellular media and apply a $30 \minute$ pulse of $50\nano \text{M}$ $\beta$-estradiol. The time-lapse microscopy movies were automatically analyzed \cite{Pelet2012} to quantify the change in nuclear localization of the GEV-mCherry protein and the fluorescent levels of the Y-Venus protein in individual cells (see S.4.5 in the Supporting Information). A calibration curve was used to map recorded intensities to total protein abundances (see S.4.3 in the Supporting Information). Based on a one-sided Kolmogorov-Smirnov test we determined and corrected for a time delay in the protein measurements, that arises from an aggregation of unmodeled sequential events, such as mRNA export, post-transcriptional/translational modifications and reporter maturation (see S.4.3 in the Supporting Information). We then used $M=20$ single cell trajectories of Y-Venus abundance and the average GEV translocation within the subsequent analyses. 

\begin{figure}[ht!]
	\centering
	\includegraphics[width=0.99\columnwidth]{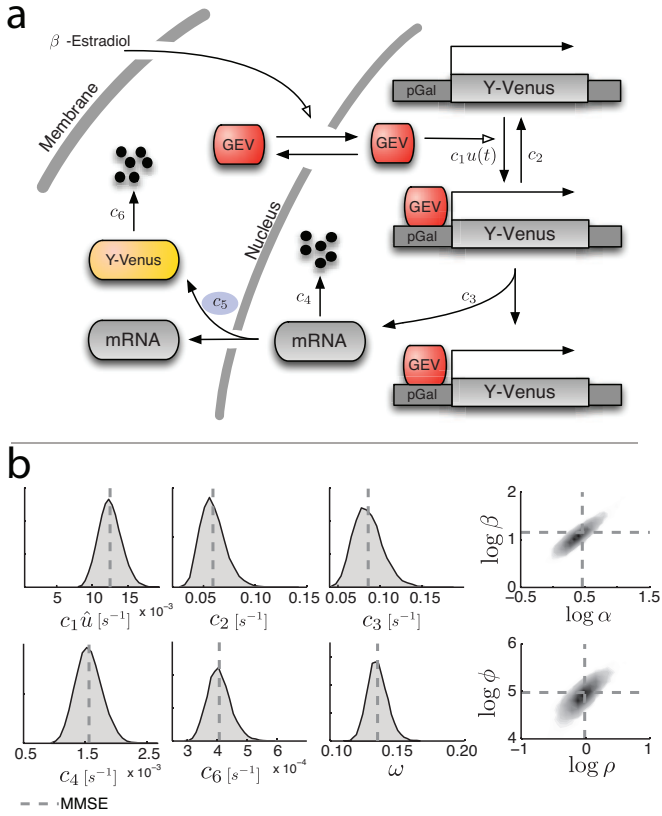}
	\caption{Parameter inference for a model of $\beta$-estradiol-induced Y-Venus expression. 
	(a) Stochastic model of $\beta$-estradiol-induced gene expression in yeast with extrinsic noise introduced by a variable translation efficiency $\kparameter_{5}$ (blue highlighted).
	(b) Posterior distributions over unknown parameters. Based on 20000 samples for the first time-iteration and 10000 samples for the subsequent iterations dynamic prior propagation was performed with respect to all kinetic parameters ($\kparameter_1,\ldots,\kparameter_4,\kparameter_6$), the acquisition noise parameter $\omega$  as well as the extrinsic statistics ($\alpha$,$\beta$) and the morphological shape parameters ($\rho$,$\phi$) that characterize the translation rate $\kparameter_{5}$. The marginal posterior for the gene-on rate is shown for $\kparameter_{1} \hat{u}$, with $\hat{u}$ as the temporal average over the modulating GEV intensity. }
		\label{fig:Figure3}
\end{figure}

\subsection*{Modeling GAL1 Y-Venus Expression and Acquisition.}
We determined empirical evidence for three models of eukaryotic gene expression by Bayesian model selection. Next to the canonical two-state model \cite{Raj2006} (see Figure 4a) we considered a model with a third refractory state \cite{Suter2011,Harper2011} (see figure S.6b in the Supporting Information) and a three-state variant of a model proposed in \cite{Blake2003}, where the initiation-complex assembly is followed by a slow activation step representing either RNA polymerase (RNAP) binding or chromatin remodeling (see figure S.6c in the Supporting Information). Bayes factors of roughly 2 and 43 dB, respectively, in favor of the two-state model were found. 
Bayes factors were also computed for two competing measurement noise models (i.e., i.i.d. normal and log-normal). Strong evidence was found for log-normally distributed measurement noise, which was reflected by a highly decisive Bayes factor ($>100$dB).

\subsection*{GAL1 Transcriptional Dynamics Exhibit Mild Bursting.}
The posterior distributions over the kinetic parameters were computed via DPP using 20000 samples during the first - and 10000 samples during the subsequent time iterations (see Figure 4b). State reconstruction and gene activity detection was performed as described previously. Figure 5a shows the reconstructed dynamics for two cells with different Y-Venus abundances. 
The order of the estimated mRNA half-life of around $10$~min and synthesis rate of $6$~molec/min are inline with previous findings \cite{Zenklusen}. The specific value of the latter is above most reported rates for constitutively expressing genes \cite{Zenklusen} which appears consistent with the fact that inducible GAL-promoter driven genes show high expression levels that can even lead to toxicity and protein-aggregation when constitutively induced \cite{Mumberg1994}. This synthesis rate together with the length of $850$~bp for the Y-Venus protein and a reported elongation speed of $2$~kb/min \cite{mason2005} for GAL-driven genes indicates that there need to be at least four RNAPs on average on the gene. The timing statistics obtained from reconstructed gene-switching paths yields that for successful initiations on average around $2.5$ transcripts per active gene state are produced, suggesting that transcription re-initiation and thus mild bursting takes place in this expression system. 

\begin{figure}[ht!]
	\centering
	\includegraphics{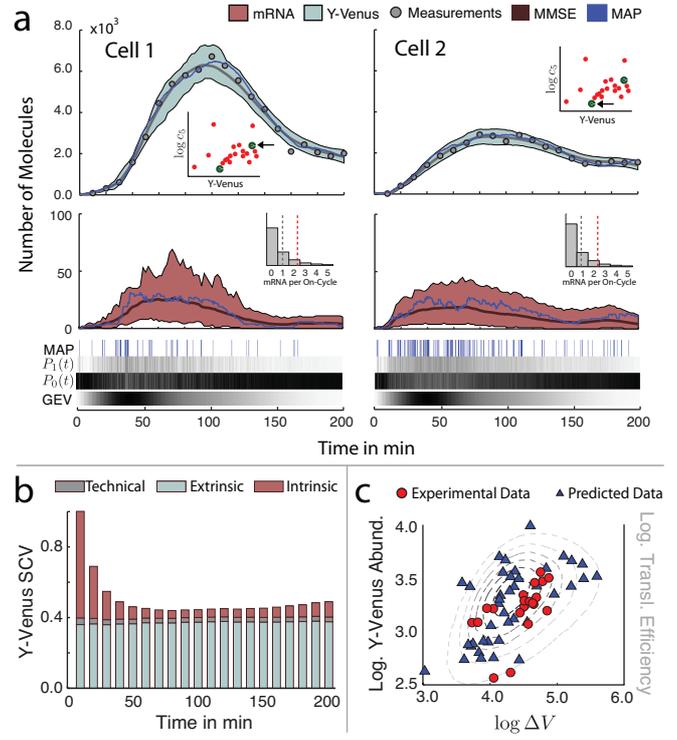}
	\caption{Reconstruction of heterogeneous Y-Venus dynamics.
	(a) Inferred dynamics for two exemplary cells showing different Y-Venus abundance. The temporal GEV-induction is shown as an intensity map, whereas white and black coloring denote minimal and maximal abundance, respectively. 
The inlet scatter plots indicate the inferred expression regime, where each dot represents a single cell and the arrows point towards the corresponding cell; the x- and y-axes correspond to the logarithm of the temporal mean of Y-Venus and the MMSE estimate of $\kparameter_{5}$, respectively. The panel below shows the inferred mRNA dynamics. Therein, the inlet with the transcripts per on cycle distribution over all posterior paths shows a mean around one (gray dashed) and mean around $2.5$ taking successful initiation events (red dashed).
	(b) Sources of cell-to-cell variability in Y-Venus expression. The inferred model was used to simulate $M=10000$ trajectories which were used to compute the squared coefficient of variation (SCV) of the Y-Venus abundance. The total SCV was decomposed into technical, intrinsic and extrinsic components (see Methods).  
	(c) Dependency between volume increase and Y-Venus abundance. Multiple cell's intensity trajectories, their volume increase and translation efficiency (i.e., $\kparameter_{5}$) were computed via forward-simulation of the inferred model. The plot shows the Y-Venus abundance at time $T=200\minute$ versus the volume increase from $T=0\minute$ until $T=200\minute$ for the predicted (blue) and experimental (red) data. 
	The statistical dependency between the volume increase and the translation efficiency is indicated by the dashed iso-lines (gray). }
		\label{fig:Figure3B}
\end{figure}

\subsection*{Noise Contributions in GAL1 Y-Venus Expression.}
Figure 5a also indicates to which extent a cell's expression level is explained by extrinsic and intrinsic factors. Although cell 1 shows mRNA levels similar to cell 2, the former expresses significantly more Y-Venus due to a larger translation rate $\kparameter_{5}$. 
By forward simulating the inferred model we dissected and quantified the different sources of variability in the measured Y-Venus abundance using the law of total variance (see Methods). In particular, we separated intrinsic, extrinsic and technical contributions to the overall variability, which are shown in Figure 5b. In the technical contribution any systematic bias introduced, for instance by the image segmentation algorithm, is not considered. The inferred model predicts that the variability in Y-Venus expression is dominantly driven by extrinsic factors. 
Although the technical contributions appear small compared to the total population variability, each intensity trajectory is characterized by significant measurement uncertainty (such as indicated in Figure 5a). 

\subsection*{Morphological Features and Extrinsic Variability.}
On top of the protein measurements, the inference procedure allows to incorporate additional single-cell readouts such as morphological features. More specifically, it is able to quantify statistical dependencies between such readouts and a population's extrinsic factors, offering an efficient way to generate and validate biological predictions. Here we hypothesized a dependency between volume increase during the observation time interval and the extrinsic factor (i.e., the translation efficiency) and quantified it using the proposed inference algorithm. The found covariation is depicted in Figure 5c. Consistent with \cite{Colman-Lerner2005} we find that volume increase correlates with translation efficiency ($r=0.56$, $p=0.01$) but that it does not explain all extrinsic variability present in the intensity trajectories. Hence, this provides more evidence for the fact that simple normalization through morphological features (e.g. forward scattering in flow cytometry data \cite{Zechner2012}) can not sufficiently correct for extrinsic variability in data.


\section*{Discussion}

Inference of stochastic dynamics of a cellular process from live-cell recordings of a single cell is inherently ill-posed. In contrast, pooling together of even a few cell recordings (e.g. around 10-20) significantly improves the accuracy of the inference and can even resolve practical non-identifiabilities (see S.3.4 in the Supporting Information). However, the large degree of extrinsic variability being evident in such data induces another layer of complexity where straightforward approaches suffer from the curse of dimensionality. 

The proposed framework rests upon a recursive inference scheme, whose strength is achieved by integrating out extrinsic factors and uncertain kinetic parameters from the process dynamics.  
First, kinetic parameters are no longer sampled along with the dynamic states resulting in a variance reduction of the desired posterior statistics (see S.3.1 in the Supporting Information). Second, in the context of cell-to-cell variability, the marginalization yields scalability with respect to the cell population size and therefore appears advantageous to previous approaches that require intermediate sampling of extrinsic factors \cite{Koeppl2012}.
The approach is conceptually related to the rao-blackwellized particle filter \cite{Doucet2000}, where a set of linear dynamic states is marginalized out in order to reduce the dimensionality. While those approaches are limited to a special class of system dynamics, DPP applies to any non-linear reaction network with linearly-parametrized propensities, such as obtained by mass-action principles. Apart from inference, the marginalized process is important in itself to properly describe the dynamics of heterogeneous cell populations. We remark that the presented inference framework is exact in the sense that for the limit of infinite samples the computed distribution converges to the correct posterior smoothing distribution. 


The simple two-state gene expression model from Figure 2a demonstrates correctness of the novel inference framework.  Although only a few measurements of the protein abundance were available, all parameters (see Figure 2b) as well as gene-switching sequences (see Figure 3) were accurately inferred from the data. 

When applied to the engineered $\beta$-estradiol-induced gene expression system in yeast, the inference scheme together with performed protein calibration measurements offers estimates in terms of absolute quantitation for kinetic parameters, unobserved molecular states and population heterogeneity. Hence, 
it allows to estimate several quantities and their uncertainty at once without resorting to dedicated experimental techniques \cite{Zenklusen, Raj2006}. The results are given for the two-state gene expression model because Bayes factors indicated weak empirical evidence for more complex models. In particular, no refractory gene state was evidenced by the data.  

By pooling heterogeneous single-cell traces we can model-based 
dissect the different contributions to cell-to-cell variability -- something that traditionally requires experiments such as two-color assays \cite{Elowitz2002,oshea}. Inline with previous studies on GAL-driven genes \cite{Blake2003,oshea} we find that extrinsic noise is the dominating source of cell-to-cell variability for such genes.  Moreover, in the course of induction we observe, consistent with Poissonian noise, a characteristic decrease of the intrinsic noise components accompanying the increase in mean expression level. 
The magnitude of technical or measurement noise was estimated to be in the same order than that of intrinsic noise.  Bayesian model selection for the measurement noise model provided evidence in favor of log-normally distributed noise with respect to the normal distribution. This indicates that measurement errors scale with the measured fluorescent intensity, such as incurred in image segmentation. Morphological features \cite{Rinott2011} in addition to intensity trajectories can significantly increase the predictive power of computational models with respect to extrinsic factors. The probabilistic dependency between translation efficiency and volume increase extracted by the algorithm is coherent with earlier findings \cite{Colman-Lerner2005} that both quantities are positively correlated.   


\section*{Acknowledgements}

We want to thank Hans Rudolf Kuensch and Jan Hasenauer for their valuable feedback on the manuscript and Odd Aalen for providing us his technical report from 1988. Furthermore, we thank Fabian Rudolf for help in designing and cloning the Y-Venus destabilized reporter and Sung-Sik Lee with the fluidic set-up. C.Z., M.U. and H.K.  acknowledge the support from the Swiss National Science Foundation, grant no. PP00P2\_128503 and SystemsX.ch, respectively. S.P. and M.P. acknowledge the support from the European project UNICELLSYS, the European Research Council, the SystemsX.ch organization (LiverX), the Competence Centre for Systems Physiology and Metabolic Disease, the Swiss National Science Foundation, and the ETH Zurich.

\section*{Author Contributions}

C.Z., S.P., M.P. and H.K. designed research. C.Z. and H.K. developed the mathematical methods, performed simulations and analyzed data. M.U. and S.P. developed strains, M.U. performed experiments and measured data. C.Z. and H.K. wrote the paper.

\section*{Methods}
\everymath{\scriptstyle}

\subsection*{Supporting Information.}  A PDF document with supplementary derivations and details is available at \href{http://www.bison.ethz.ch/bison/SI_Pooling/}{http://www.bison.ethz.ch/bison/SI\textunderscore Pooling/}.

\subsection*{Experimental Procedures.}
Measurements were performed in an incubation chamber at $30 \celsius$ using an automated epi-fluorescence microscope (Eclipse Ti, Nikon Instruments) with a $60 \mathrm{x}$ oil objective and specific (CFP/YFP/mCherry) excitation and emission filters. Images were acquired at multiple positions using a motorized XY-stage, while the focal plane was maintained using a hardware autofocus system. Cell chambers ($\mu \text{-Slide VI}^{0.4}$, ibidi) were treated with filtered solution of Concanavalin A dissolved in PBS ($1 \milli\gram /  \milli\litre $) for $30 \minute$ and subsequently rinsed with PBS. Single colonies of the respective yeast strain were picked and inoculated in synthetic (SD) medium. The overnight saturated cultures were then diluted and grown in log-phase for at least two doubling times ($ > 4 \hour$), again diluted (OD600 0.01), briefly sonicated and loaded into the cell chambers. Pulse experiments were performed by switching between two hydrostatic pressure driven liquid flows (see section S.4 in the Supporting Information). A calibration curve was obtained by measuring fluorescence intensities for several proteins of known average abundance, tagged with a Venus fluorescent reporter. For a graphical illustration and further details see S.4.3 in the Supporting Information.

\subsection*{Inference Algorithm.}
A full technical description of the sampling-based inference algorithm can be found in section S.2 in the Supporting Information. Empirical model evidences - and Bayes factors are directly calculated by the algorithm and do not require further computations. The corresponding formulas and their derivations are given in S.2.6 in the Supporting Information. A corresponding Matlab toolbox with a simple user interface is made available under the GNU public license.  

\subsection*{Model-Based Noise Decomposition.}
The law of total variance is applied to dissect the total variability into intrinsic, extrinsic and technical contributions. In particular, it holds that
\begin{equation}
\begin{split}
	&\mathrm{SCV}[\Measurement_{l}] =  \underbrace{\frac{\Expect{\Expect{\Var{\Measurement_{l} \mid \State_{l}}\mid \ExtrinsicVariable}}}{\Expect{\Measurement_{l}}^{2}}}_{\text{technical}} \\
	&\quad  + \underbrace{\frac{\Expect{\Var{\Expect{\Measurement_{l} \mid \State_{l}} \mid \ExtrinsicVariable}}}{\Expect{\Measurement_{l}}^{2}} }_{\text{intrinsic}} +\underbrace{\frac{\Var{\Expect{\Expect{\Measurement_{l}\mid \State_{l}}\mid \ExtrinsicVariable}}}{\Expect{\Measurement_{l}}^{2}}}_{\text{extrinsic}}.
\end{split} \nonumber
\end{equation}
For the decomposition, the model parameters were set to their inferred MMSE estimates and the individual quantities were obtained via forward simulation. Further details can be found in section S.2.7 in the Supporting Information.

{\footnotesize \input{Main.bbl}}

\end{document}

%% file: Definitions.tex
\def\TimeIndex{N}
\def\timeIndex{l}

\def\extrinsicVariable{z}
\def\ExtrinsicVariable{Z}

\def\hyperParameters{a}
\def\HyperParameters{A}

\def\MarkovChain{\State}
\def\State{X}

\def\StateInt{\mathbf{\State}}
\def\state{x}
\def\stateInt{\mathbf{\state}}

\def\stateDiscreteTime{\state_{\timeIndex}}
\def\StateDim{d}
\def\ReactionDim{\nu}
\def\CellCount{M}

\def\Measurement{Y}
\def\MeasurementInt{{\Measurement}}
\def\measurement{y}
\def\measurementInt{{\measurement}}

\def\measurementTime{\measurement_{\timeIndex}}

\def\AllParameters{\Theta}

\def\SharedParameters{S}
\def\sharedParameters{s}

\def\kParameters{C}
\def\kparameters{c}
\def\kparameter{c}

\def\MargProp{h}

\def\MeasurementParameters{\Omega}
\def\measurementParameters{\omega}

\def\Covariates{V}
\def\covariates{v}

\def\GammaDist{\mathcal{G}}

\def\df{\mathrm{d}}

\newcommand{\Expect}[1]{ \mathbb{ E } \left[ #1 \right]}

\newcommand{\Var}[1]{ \mathrm{Var} \left[ #1 \right]}

\def\MorphParameters{B}
\def\morphParameters{b}